\documentclass[twocolumn]{aastex631}

\usepackage{xspace}	    

\begin{document}
\include{macros}

\title{An extremely compact, low-mass post-starburst galaxy at $z=5.2$}

\correspondingauthor{Victoria Strait}
\email{victoria.strait@nbi.ku.dk}


\author[0000-0002-6338-7295]{Victoria Strait}
\affiliation{Cosmic Dawn Center (DAWN), Denmark}
\affiliation{Niels Bohr Institute, University of Copenhagen, Jagtvej 128, DK-2200 Copenhagen N, Denmark}

\author[0000-0003-2680-005X]{Gabriel Brammer}
\affiliation{Cosmic Dawn Center (DAWN), Denmark}
\affiliation{Niels Bohr Institute, University of Copenhagen, Jagtvej 128, DK-2200 Copenhagen N, Denmark}

\author[0000-0002-9330-9108]{Adam Muzzin}
\affiliation{Department of Physics and Astronomy, York University, 4700 Keele St., Toronto, Ontario, Canada, MJ3 1P3}

\author[0000-0001-8325-1742]{Guillaume Desprez}
\affiliation{Institute for Computational Astrophysics and Department of Astronomy \& Physics, Saint Mary’s University, 923 Robie Street, Halifax, NS B3H 3C3, Canada}

\author[0000-0003-3983-5438]{Yoshihisa Asada}
\affiliation{Institute for Computational Astrophysics and Department of Astronomy \& Physics, Saint Mary’s University, 923 Robie Street, Halifax, NS B3H 3C3, Canada}
\affiliation{Department of Astronomy, Kyoto University, Sakyo-ku, Kyoto 606-8502, Japan}

\author[0000-0002-4542-921X]{Roberto Abraham}
\affiliation{Dunlap Institute for Astronomy and Astrophysics, 50 St. George Street, Toronto, Ontario, M5S 3H4, Canada}

\author[0000-0001-5984-0395]{Maru{\v s}a Brada{\v c}}
\affiliation{University of Ljubljana, Department of Mathematics and
Physics, Jadranska ulica 19, SI-1000 Ljubljana, Slovenia}

\author[0000-0001-9298-3523]{Kartheik G. Iyer}
\altaffiliation{Hubble Fellow}
\affiliation{Columbia Astrophysics Laboratory, Columbia University, 550 West 120th Street, New York, NY 10027, USA}

\author[0000-0003-3243-9969]{Nicholas Martis}
\affiliation{NRC Herzberg, 5071 West Saanich Rd, Victoria, BC V9E 2E7, Canada}
\affiliation{Institute for Computational Astrophysics and Department of Astronomy \& Physics, Saint Mary’s University, 923 Robie Street, Halifax, NS B3H 3C3, Canada}

\author[0000-0002-8530-9765]{Lamiya Mowla}
\affiliation{Dunlap Institute for Astronomy and Astrophysics, 50 St. George Street, Toronto, Ontario, M5S 3H4, Canada}

\author{Gaël Noirot}
\affiliation{Institute for Computational Astrophysics and Department of Astronomy \& Physics, Saint Mary’s University, 923 Robie Street, Halifax, NS B3H 3C3, Canada}

\author[0000-0001-8830-2166]{Ghassan Sarrouh}
\affiliation{Department of Physics and Astronomy, York University, 4700 Keele St., Toronto, Ontario, Canada, MJ3 1P3}

\author[0000-0002-7712-7857]{Marcin Sawicki}
\altaffiliation{Canada Research Chair}
\affiliation{Institute for Computational Astrophysics and Department of Astronomy \& Physics, Saint Mary’s University, 923 Robie Street, Halifax, NS B3H 3C3, Canada}

\author[0000-0002-4201-7367]{Chris Willott}
\affiliation{NRC Herzberg, 5071 West Saanich Rd, Victoria, BC V9E 2E7, Canada}

\author[0000-0003-4196-5960]{Katriona Gould}
\affiliation{Cosmic Dawn Center (DAWN), Denmark}
\affiliation{Niels Bohr Institute, University of Copenhagen, Jagtvej 128, DK-2200 Copenhagen N, Denmark}

\author[0009-0006-4881-3299]{Tess Grindlay}
\affiliation{NRC Herzberg, 5071 West Saanich Rd, Victoria, BC V9E 2E7, Canada}
\affiliation{Department of Physics and Astronomy, University of Victoria, Victoria, BC, V8P 5C2, Canada}

\author[0000-0002-7547-3385]{Jasleen Matharu}
\affiliation{Cosmic Dawn Center (DAWN), Denmark}
\affiliation{Niels Bohr Institute, University of Copenhagen, Jagtvej 128, DK-2200 Copenhagen N, Denmark}

\author[0009-0009-4388-898X]{Gregor Rihtar{\v s}i{\v c}}
\affiliation{University of Ljubljana, Department of Mathematics and
Physics, Jadranska ulica 19, SI-1000 Ljubljana, Slovenia}

\begin{abstract}
We report the discovery of a low-mass $z=5.200\pm 0.002$ galaxy that is in the process of ceasing its star formation. The galaxy, MACS0417-z5PSB, is multiply imaged with magnification factors $\sim40$ by the galaxy cluster MACS J0417.5-1154, observed as part of the CAnadian NIRISS Unbiased Cluster Survey (CANUCS). Using observations of MACS0417-z5PSB with a JWST/NIRSpec Prism spectrum and NIRCam imaging, we investigate the mechanism responsible for the cessation of star formation of the galaxy, and speculate about possibilities for its future. 
Using spectrophotometric fitting, we find a remarkably low stellar mass of $\rm{M_*}=4.3\pm^{0.9}_{0.8} \times 10^{7}  \rm{M_{\odot}}$, less than 1\% of the characteristic stellar mass at $z\sim5$. We measure a de-lensed rest-UV half-light radius in the source plane of $30\pm^{7}_{5}$ pc, and measure a star formation rate from H$\alpha$ of $0.14\pm^{0.17}_{0.12}$ $\rm{M_{\odot}/yr}$. 
We find that under the assumption of a double power law star formation history, MACS0417-z5PSB has seen a recent rise in star formation, peaking $\sim10-30$ Myr ago and declining precipitously since then. Together, these measurements reveal a low-mass, extremely compact galaxy which is in the process of ceasing star formation. We investigate the possibilities of mechanisms that have led to the cessation of star formation in MACS0417-z5PSB, considering stellar and AGN feedback, and environmental processes. We can likely rule out an AGN and most environmental processes, but leave open the possibility that MACS0417-z5PSB could be a star forming galaxy in the lull of a bursty star formation history. 

\end{abstract}


\section{Introduction} \label{sec:intro}

Some of the most intriguing open questions in galaxy evolution include how and when galaxies start and cease forming their stars, and subsequently, how those quiescent galaxies continue to grow. To answer these questions, populations of quiescent galaxies have been observed at earlier and earlier epochs over the years, as early observations are necessary to distinguish between models (e.g., \citealp{lovell2022, Wellons2023}), as well as to settle tensions between models and observations \citep{Merlin2019, Santini2021, Gould2023}. 
Multiple scenarios have been proposed to explain the growth of quiescent galaxies from $z\sim3$ to the local universe. The two most likely dominant scenarios are either the growth of star-forming galaxies via in situ star formation and gas accretion, followed by a ``monolithic collapse", or a series of dry mergers of small, early quiescent galaxies (e.g., \citealp{Bezanson2009, Naab2009, Hopkins2010}). 

These scenarios are likely intimately related to the quenching mechanisms of the galaxies themselves, which can occur on a variety of timescales (see e.g., \citealp{Man2018, Iyer2020, Sherman2020, Tacchella2022, Noirot2022a}). These quenching mechanisms likely depend on galaxy and environment properties--e.g., for a low-mass satellite galaxy, environmental quenching is expected to play a larger role via overconsumption \citep{McGee2014}, starvation \citep{Whitaker2021, Williams2021}, or gas removal by a nearby AGN \citep{Fabian2012, King2015}. The effects of these processes over various eras in the universe have been investigated by simulations (e.g., \citealp{Trayford2015, Nelson2018, Nelson2019, Donnari2021, Wu2021, Ward2022}) and require more observations at early times to distinguish possibilities. 

These debates have led to great interest in observations of quiescent galaxies, increasingly early in the universe. However, possibly due to observational constraints, the majority of the literature about quiescent galaxies has focused on low-redshift, high-mass sources (with some exceptions, e.g., \citealp{Santini2022, Weaver2022, Looser2023}). The highest redshift quiescent/post starburst galaxies to date are at $z=7.3$ \citep{Looser2023}, followed by $z=4.6$ \citep{Carnall2023a}, followed then by several quiescent galaxies at $z\sim2-4$ \citep{Glazebrook2017, Schreiber2018, Forrest2020, Valentino2020, Marchesini2023, Nanayakkara2022}, most with stellar masses $\gtrsim10^{10}\rm{M_{\odot}}$.

In this work, we investigate a multiply-imaged post-starburst galaxy at $z=5.200\pm 0.002$ behind galaxy cluster MACS J0417.5-1154, first discovered by \cite{Mahler2019}, which we dub MACS0417-z5PSB. Each of the two images is magnified by a factor of $\sim40$, and measure de-magnified stellar mass for the main bulge of the galaxy to be only $\rm{M_*}=4.3\pm^{0.9}_{0.8} \times 10^{7}  \rm{M_{\odot}}$, less than 1\% of the characteristic stellar mass at $z\sim5$. This source allows us to investigate in detail the quenching of star formation in a low-mass galaxy at high redshift. Using spectrophotometric fitting and size measurement, we investigate possibilities for the source's past and future, and discuss the implications for the formation of massive quiescent galaxies later in the universe's history. 

This paper is structured as follows. In Section~\ref{sec:data} we describe the data used in our work. In Section~\ref{sec:methods} we describe the methods used for constraining stellar properties and size. We describe our results and compare to the literature in Section~\ref{sec:res}. We discuss our results in Section~\ref{sec:disc} and state our conclusions in Section~\ref{sec:conclusions}.

We assume a flat $\Lambda\mathrm{CDM}$ cosmology with $\Omega_m=0.3,\,\Omega_\Lambda=0.7,\,h=0.7$, all magnitudes are in the AB system, and all distances are proper unless specified otherwise.
\begin{figure*}
    \centering
    \includegraphics[width=18cm]{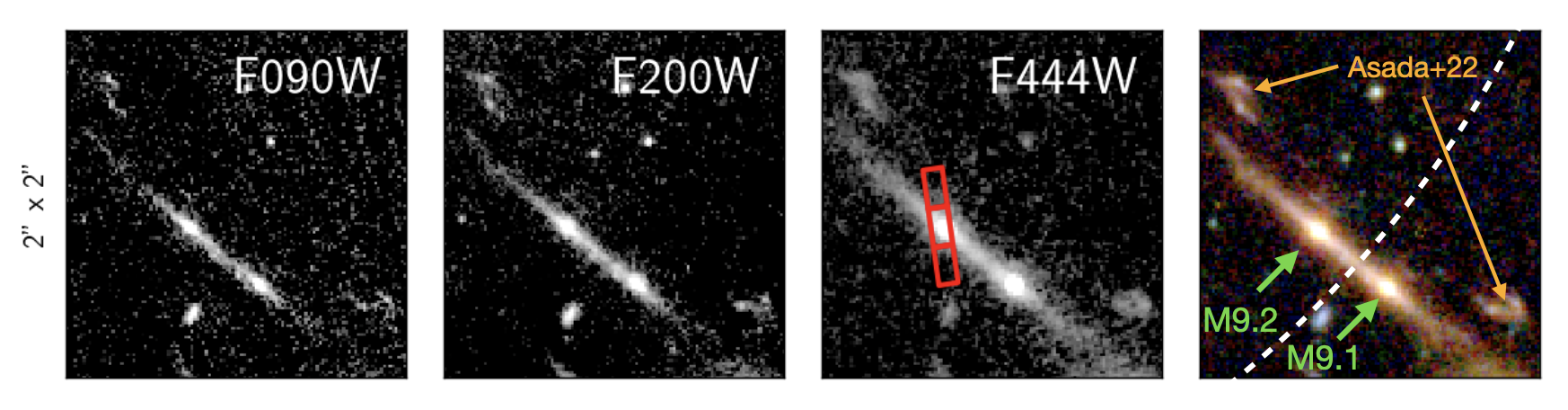}
    \caption{JWST/NIRCam images of MACS0417-z5PSB and nearby system from \cite{Asada2022}. From left to right: F090W, F200W, F444W, and RGB (red: F356W+F410M+F444W, green: F200W+F277W+F356W, blue: F090W+F115W+F150W) image of MACS0417-z5PSB. All images have 40mas pixels, are non-PSF-homogenized. FOV is $2'' \times 2''$. NIRSpec slit configuration is shown in the F444W image. Approximate critical curve, as well as the multiple images of MACS0417-z5PSB (labeled M9.1 and M9.2) and galaxies from \cite{Asada2022} are shown in the RGB image.}
    \label{fig:rgb}
\end{figure*}
\section{Data} \label{sec:data}
In this work, we utilize data from the MACS J0417.5-1154 cluster field obtained via the Canadian NIRISS Unbiased Cluster Survey (CANUCS, \citealp{Willott2022}). We use JWST/NIRSpec Prism data for spectro-photometric fitting to obtain stellar properties of the galaxy, and use all available NIRCam imaging for photometric calibration, and NIRCam SW imaging to measure the rest-UV size of the galaxy.

\subsection{NIRCam Imaging}\label{sec:imaging}
The cluster was observed with JWST/NIRCam filters F090W, F115W, F150W, F200W, F277W, F356W, F410M and F444W with exposure times of 6.4 ks each, reaching S/N between 5 and 10 for a AB=29 point source. We also utilize archival HST imaging in F105W, F125W, F140W, F160W, F435W, F606W and F814W in the SED fitting. 

The target galaxy was identified in NIRCam images as being multiply-imaged and at a redshift of around 5 based on the photometric redshift and the lensing geometry. To reduce the imaging and extract photometry, we closely follow the procedure outlined by \cite{Noirot2022b}. However, for this source we do custom photometry of $0.3''$-diameter aperture size for each clump of the galaxy on the F444W PSF-matched images, similar to the $0.2''$ width of the NIRSpec slit, making this aperture an appropriate one for flux calibration of the bulge of one of the clumps. This is, however, not all of the light from the galaxy-- we focus on the stellar properties of the bulge of the galaxy, which is $\sim75$\% of the light of the entire galaxy. The source is $\sim25.2$ mag in F444W. In Figure \ref{fig:rgb}, we show cutouts of F090W, F200W, F444W, and an RGB image incorporating all NIRCam bands of both MACS0417-z5PSB and a neighboring galaxy at the same redshift. In the F444W image, we overplot the NIRSpec slit, and in the RGB image, we plot the approximate critical curve from the best-fit lens model.

\subsection{NIRSpec Spectroscopy}\label{sec:spec}
We observed one image of the galaxy (labeled M9.2 in Figure \ref{fig:rgb}) with the JWST/NIRSpec micro-shutter assembly in the low resolution Prism mode for 2900 seconds, split over three nodded exposures. 
To reduce the spectroscopy, we closely follow the procedure outlined by \cite{Morishita2022b}. Briefly, we use the JWST pipeline\footnote{https://github.com/spacetelescope/jwst} for the Level 1 data products and \texttt{msaexp}\footnote{https://github.com/gbrammer/msaexp} for Levels 2 and 3 products. We perform background subtraction by subtracting the background level estimated from the nodded shutters in adjacent exposures.  We then use \texttt{msaexp} to extract the 1D spectrum using an inverse-variance weighted kernel following \cite{Horne1986}. We photometrically calibrate the data to account for wavelength-dependent slit losses by fitting a polynomial to the ratio of the HST+NIRCam photometry and the spectrum integrated through each filter bandpass and then multiplying the spectrum by the output polynomial. This is done as part of the spectrophotometric fitting, described in Section \ref{sec:bagpipes}.

\subsection{Gravitational lensing models}  \label{sec:data_lens}
Our work focuses on a multiply imaged galaxy behind a galaxy cluster, so to measure its intrinsic properties we use a model of the cluster's mass and magnification distribution. We use a lens model of MACS0417 (Desprez 2023, in prep.) created with \texttt{Lenstool} (\citealp{Kneib1993, Jullo2007}), starting from the parameters described by \cite{Mahler2019} but improved with several new spectroscopically confirmed multiply imaged systems, including MACS0417-z5PSB and a nearby extreme line emitter at the same redshift \citep{Asada2022}. We refer to the individual images in MACS0417-z5PSB as M9.1 and M9.2, as is done in Desprez 2023, in prep. Because the source is so close to a critical curve, differential magnification could be at play. For this reason, we focus mainly on the bulge of the galaxy which is compact, and assume only one magnification value.

We use \texttt{Lenstruction} \citep{Yang2020} to measure the size of the galaxy, which makes necessary local corrections to the lens model. This is described in more detail in Section \ref{sec:size}. To obtain uncertainties on magnification factor, we run \texttt{Lenstruction} 100 times, using the range of models produced in the MCMC chain with \texttt{Lenstool} and report the 68\% confidence limits.

\section{Methods}  \label{sec:methods}

\begin{deluxetable}{ll}[h!]
\tabletypesize{\normalsize}
\tablecaption{\label{tbl:props} } 
\tablewidth{10cm}
\tablehead{
\colhead{Property} &  \colhead{Result} 
}
\startdata
$z$ & $5.200\pm 0.002$\\
$\mu$ & $43\pm^{10}_{7}$ \\
log10($M_*/M_{\odot}$)&  $7.63\pm^{0.08}_{0.09}$ \\
A$_{\rm V}$ & $0.19\pm^{0.04}_{0.02}$\\
logU &  $-3.7\pm^{0.4}_{0.2}$\\
Age (Mass-weighted, Myr) & $37.9\pm^{4.2}_{5.6}$ \\
Stellar metallicity ($Z_{\odot}$) & $1.0\pm^{0.01}_{0.02}$ \\
SFR$_{\rm{H}\alpha}$ (M$_{\odot}/$yr) & $0.14\pm^{0.17}_{0.12}$ \\
Half-light radius (pc) & $30\pm^{7}_{5}$\\
UV $\beta$ slope & $-1.61 \pm 0.05$\\
\enddata
\tablenotetext{}{All values from SED fitting are quoted using statistical and magnification errors only (where relevant), under assumptions described in Section \ref{sec:bagpipes}. All relevant values have been corrected for magnification.}
\end{deluxetable}

\subsection{Spectro-photometric Fitting}\label{sec:bagpipes}
To model the stellar properties of MACS0417-z5PSB (e.g., stellar mass, star formation history), we use the spectrophotometric fitting code BAGPIPES \citep{Carnall2018}. Because the NIRSpec/Prism observing mode has a variable spectral resolution, we implement wavelength sampling to match the spectral resolution of the prism into the BAGPIPES fitting routine to account for this. We assume a double power law star formation history, which is commonly assumed for quiescent and post-starburst galaxies (e.g., \citealp{Carnall2018}), allowing the slopes of each side to vary freely. We assume the \cite{Charlot2000} dust attentuation recipe, allow ionization parameter logU to vary between -4 and -1, and assume a Kroupa IMF \citep{Kroupa2001}. All other parameters are left free.
\subsection{Size Measurement} \label{sec:size}
To measure the size of MACS0417-z5PSB, we utilize the public code \texttt{Lenstruction}\footnote{https://github.com/ylilan/lenstruction} \citep{Yang2020}. We use a version of the NIRCam F200W image drizzled onto a 20mas pixel scale to measure the rest-frame UV size, and choose a nearby star to model the point spread function. We use the interactive mode of \texttt{Lenstruction}, allowing us to select for any nearby contaminating light or lens light in the image. We assume a S\'ersic model for the galaxy's morphology, and allow S\'ersic index and ellipticity to vary freely.

\section{Results} \label{sec:res}
\subsection{Stellar properties}
\begin{figure*}
    \centering
    \includegraphics[width=17cm]{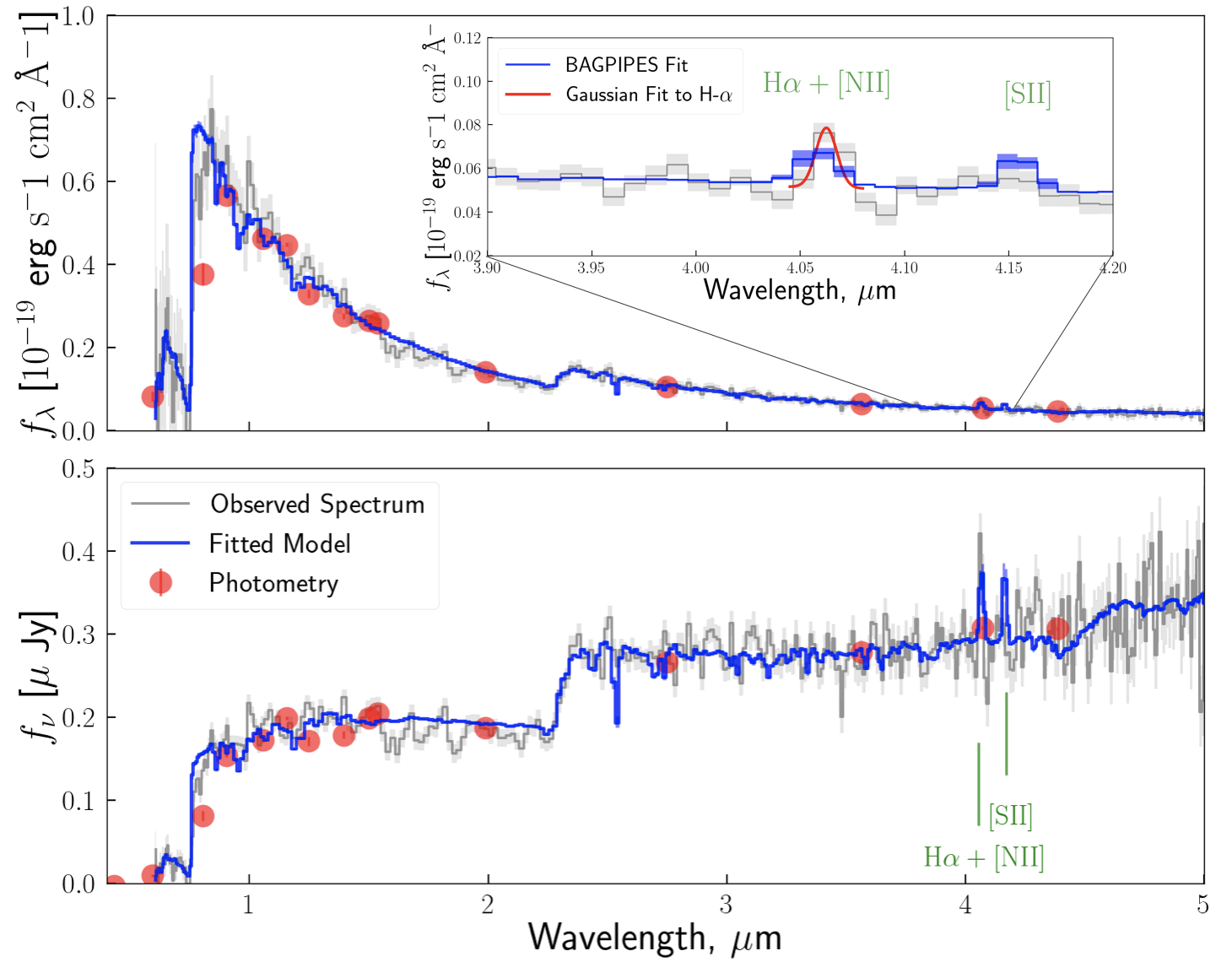}
    \caption{Photometrically calibrated observed spectrum (dark gray) with 1-$\sigma$ noise (light gray), and posterior spectrum from BAGPIPES (blue) for MACS0417-z5PSB. Top panel in $f_{\lambda}$, bottom panel in $f_{\nu}$. Emission lines labeled in green. Inset is a zoomed-in version of the spectrum surrounding the detection of H$\alpha$. Note that although BAGPIPES fits an [SII] line, we do not detect it in our spectrum.}
    \label{fig:spec}
\end{figure*}

In Figure \ref{fig:spec} we show the photometrically calibrated, observed spectrum and 1-$\sigma$ noise in gray. Lyman and Balmer breaks are clearly identified at $\sim7540$ and $\sim23500$ \AA, and a detection of H$\alpha$ at $\sim40700$ \AA. To obtain a redshift, we fit a Gaussian profile to the H$\alpha$ line at 40700 \AA, we find a redshift of $5.200\pm0.0002$. 

In the inset panel of Figure \ref{fig:spec}, we show a zoomed in version on the H$\alpha$ detection. While there is a fit from BAGPIPES, we infer properties of H$\alpha$ via a separate fit of a Gaussian line to the relevant spectral region, shown in red. H$\alpha$ and [NII] are blended in the prism spectra, so following \cite{Wuyts2013}, we assume that [NII] contibutes 15\% of the H$\alpha$+[NII] flux. After correcting for contribution from [NII], we find that the SFR (corrected for magnification, and including magnification uncertainties) is $0.14\pm^{0.17}_{0.12} \rm{M_{\odot}}/ \rm{yr}$. In the 2D spectrum, the H$\alpha$ line, while partially coincident with the rest of the spectrum, extends farther than the continuum, suggesting that some of the emission is coming from the disk of the galaxy, rather than the bulge which we mainly focus on here.

\begin{figure*}[ht!!!]
    \centering
    \includegraphics[width=18cm]{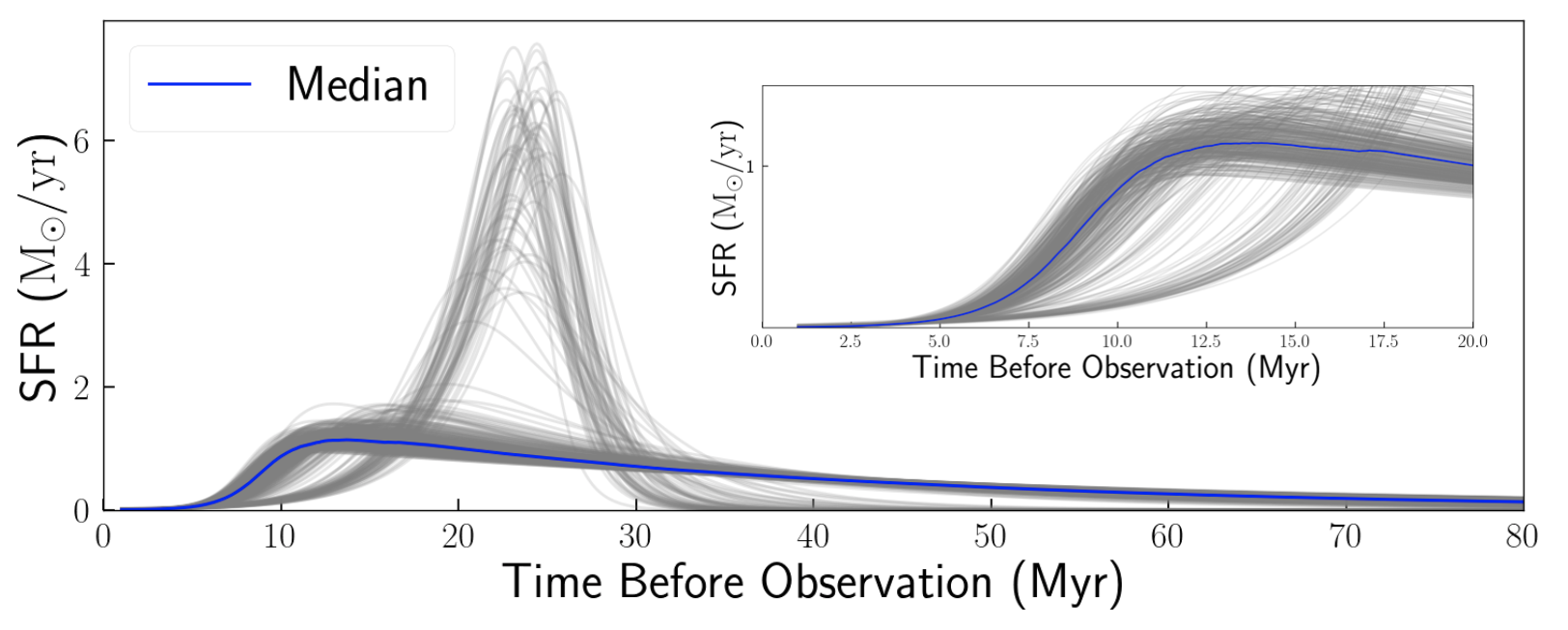}
    \caption{Star formation history posteriors from BAGPIPES. Median star formation history in solid blue line, all other possibilities in gray. Inset is a zoomed in version of the same plot, $<20$ Myr before observation. MACS0417-z5PSB peaked in star formation $\sim10-30$ Myr ago and the SFR has declined precipitously since then. Median magnification factor of 43 assumed.}
    \label{fig:sfh}
\end{figure*}
Figure \ref{fig:spec} also shows the posterior spectrum from our BAGPIPES run in blue, and in Table \ref{tbl:props} we list medians and 1-$\sigma$ uncertainties of the posteriors for select stellar properties of the galaxy. We find a stellar mass of $\rm{log(M_*/M_{\odot})} = 7.63\pm^{0.08}_{0.09}$, accounting for a magnification correction factor of $\mu=43\pm^{10}_{7}$, $<1$\% of the characteristic stellar mass at $z\sim5$ \citep{Weaver2022}.  This places MACS0417-z5PSB within an extrapolation of the star forming main sequence at $z\sim5$ \citep{Salmon2015}. The star formation rate averaged over the last 10 Myr from BAGPIPES is $0.10\pm_{.05}^{.06} \rm{M_{\odot}}/ \rm{yr}$.

In Figure \ref{fig:sfh}, we show the posterior star formation history for the galaxy, revealing either a steeply or more slowly rising burst of star formation, followed by a precipitous decline, $\sim10-30$ Myr before observation. We find that the shape of the star formation history is mainly driven by the Lyman and Balmer breaks, and is not heavily affected by resolution effects in the spectrum.

\subsection{Size}\label{sec:sizeresults}
\begin{figure*}
    \centering
    \includegraphics[width=16cm]{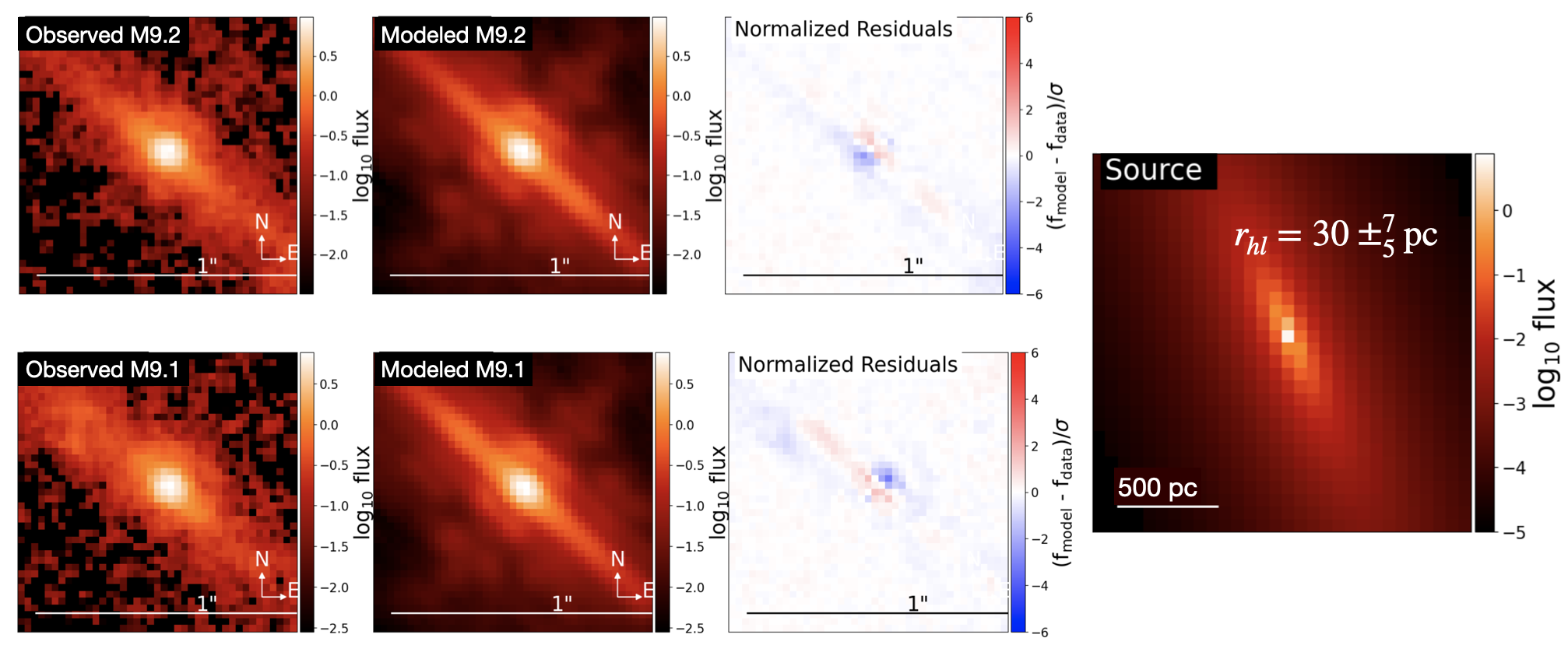}
    \caption{\textbf{Top left:} One observed multiple image of MACS0417-z5PSB, M9.2, followed by the model and normalized residuals from Lenstruction. The hexagonal structure in the images comes from the JWST PSF. \textbf{Bottom left:} The same, for the second image, M9.1. \textbf{Right:} Modelled source plane image.} 
    \label{fig:size}
\end{figure*}

In Figure \ref{fig:size}, we show the results from our rest-UV size measurement with \texttt{Lenstruction}. From left to right, we show the observed, modelled, and residuals of each source, and finally the modelled image in the source plane. We find a remarkably small size of $r_{hl} = 30\pm^{7}_{5}$ pc for MACS0417-z5PSB, where $r_{hl}$ is half-light radius. While new results on rest-UV sizes at $z\sim5$ with JWST are sparse, this size is much smaller than average sizes reported for galaxies at $z\gtrsim7$ (e.g., \citealp{Yang2022, Ono2022}), but similar to some of the single-object discoveries found so far (e.g., \citealp{Williams2022}). 

\section{Discussion} \label{sec:disc}
\subsection{The galaxy's past: why did star formation cease?}
The star formation history of MACS0417-z5PSB in Figure \ref{fig:sfh} reveals a recent rise and sharp decline in star formation, less than 30 Myr before observation. We briefly investigate three options for quenching possibilities in this galaxy: stellar and AGN feedback, exhaustion, and environmental quenching. We also discuss the possibility that the source is a star forming galaxy in a lull of a bursty star formation history.

\subsubsection{AGN and Stellar Feedback}
There is growing evidence that low-mass ($\sim10^8$ $\rm{M_{\odot}}$) galaxies are commonly hosts to AGN, both from observations (e.g., \citealp{ManzanoKing2019, Ding2020}) and cosmological zoom-in simulations (e.g., \citealp{Sharma2022, Koudmani2022, Reines2022}). In particular, \cite{Almeida2023} find that low-luminosity AGN are capable of quenching galaxies, although this study only explores down to AGN of $10^6$ $\rm{M_{\odot}}$, a higher mass than would be expected for an AGN in MACS0417-z5PSB. While inconclusive, this hints toward the possibility that feedback from AGN could be contributing to the cessation of star formation in MACS0417-z5PSB. While it is possible that this effect in combination with stellar feedback (e.g., from supernovae and stellar winds) which would be expected in low mass galaxies, could have caused the cessation of star formation in this galaxy, we do not see evidence of an AGN via a a broad component from H$\alpha$ and can likely rule out a scenario where an AGN is the dominant quenching mechanism.

\subsubsection{Exhaustion}
The small size and high stellar density of the galaxy suggests the possibility that the galaxy simply formed its stars very rapidly and exhausted its gas. Studies such as \cite{Wu2018, Belli2019} show that post-starburst galaxies tend to be smaller than their quiescent counterparts, and show evidence for a ``faster track" to quiescence for galaxies, pointing towards a violent event like a major merger, which could have possibly spurred star formation. The high inferred stellar metallicity from BAGPIPES (see Table \ref{tbl:props}) may also indicate a preference for this solution-- if there were an outflow from an AGN and/or stellar feedback, one would expect metals to have been ejected in the outflow. 

\subsubsection{Environmental factors}
It has been shown that environmental quenching is the dominant quenching mechanism at low-$z$ \citep{Peng2010}. While large enough groups and clusters where environmental quenching factors would come into play are not likely to have formed as early as $z\sim5$, it may be possible that there are small overdensities which could contribute to this galaxy's current state, albeit with less dense and violent surroundings. It is worth noting that there is a nearby system \citep{Asada2022} at the same redshift as MACS0417-z5PSB. It is $\sim2.5$ kpc away from MACS0417-z5PSB in the source plane, has a stellar mass of $\sim10^7$ $\rm{M_{\odot}}$, and is in the process of a merger and undergoing a burst of star formation. While this is a low-mass system, it may point to the presence of a higher mass system nearby which could contribute to environmental quenching. However, we do not see evidence for a high-mass system at the same redshift in the JWST FOV via a selection using photometric redshifts, so can likely rule out many environmental mechanisms as the dominant force at play in MACS0417-z5PSB.

\subsection{The galaxy's future: will it rejuvenate?}
MACS0417-z5PSB appears to be in the process of turning off its star formation, as evidenced by a lack of emission lines, but with no sign of Balmer absorption lines yet (although their presence cannot be ruled out with the low-resolution Prism spectrum). We cannot constrain the age of the stars to higher precision without the resolved Balmer absorption lines, but we can see that the galaxy is not currently undergoing significant star formation. Combined with its remarkably small size and high inferred stellar density, this points to the possibility of the galaxy being a progenitor of quiescent galaxies observed later in the history of the universe, and may point to early quenching and dry merging as a method of future growth, following the path of quiescent galaxy growth in, e.g.,  \cite{vanderWel2014}.

However, it is difficult to rule out the possibility that MACS0417-z5PSB is a star-forming galaxy in a lull of a stochastic star formation history, with a future of rejuvenation via a wet merger or gas accretion. The \texttt{Astraeus} Simulations \citep{legrand2022} find that galaxies are likely to form most of their mass in a stochastic star formation phase at stellar masses at or below $\sim10^7$ $\rm{M_{\odot}}$.  Similarly in the FIRE simulations, \cite{Ma2019} find that $10^8$ $\rm{M_{\odot}}$ is the mass threshold where star formation histories begin to transition from bursty to stable. Because MACS0417-z5PSB resides on the edge of this limit, it is hard to say how likely a bursty star formation is. It is possible that the stellar mass of this galaxy is too high for a likely stochastic star formation history and chance of rejuvenation. However, if the quenching mechanism is not environmental, the stellar mass ($\sim 10^{7.63}$ $\rm{M_{\odot}}$) may be too low to fully, permanently quench: most observed high-$z$ galaxies to date have had stellar masses above $10^{10}$ $\rm{M_{\odot}}$, and most investigations of passive galaxies in simulations are for log$(\rm{M_*/M_{\odot}})<10^{9} $. It may be that a deep gravitational potential from a larger dark matter halo is required to heat any remaining gas in a passive galaxy in order to permanently quench, and that MACS0417-z5PSB will be likely to rejuvenate from infalling gas or wet mergers. 

\subsection{Implications for Compaction}
It is a long-standing idea that surface density within 1 kpc of a galaxy's center is a good indicator of whether or not a galaxy has been quenched (e.g., \citealp{Zolotov2015, Tacchella2016}). While this may be true for galaxies of log10($\rm{M_*}$)$\gtrsim9$, this indicator would not be appropriate for MACS0417-z5PSB, even if it does continue to cease star formation and quench, as the total radius is far below 1 kpc. While MACS0417-z5PSB does not meet the sSFR criterion to be called quenched, it is one of the first of its kind at such an early time in the universe, and it seems that a new ``compaction" indicator for galaxies at high redshifts would be called for; for example, the stellar density within 0.1 kpc instead of 1 kpc using the definition in \cite{Zolotov2015} would have selected MACS0417-z5PSB as a candidate quenched galaxy.


\section{Conclusions and Future Data}\label{sec:conclusions}

Our conclusions are as follows:

\begin{enumerate}
    \item We report the discovery of a post-starburst, nearly quiescent galaxy at $z=5.200\pm 0.002$. For the bulge of the galaxy, we measure a star formation rate of $0.14\pm^{0.17}_{0.12} \rm{M_{\odot}}/ \rm{yr}$, and a stellar mass of $\rm{M_*}=4.3\pm^{0.9}_{0.8} \times 10^{7}  \rm{M_{\odot}}$. 
    
    \item We find that the galaxy's star formation history reveals a recent rise in star formation, formed $\sim10-30$ Myr ago and follwed by a steep decline. Combined with the presence of a nearby system at the same redshift, this may give hints about its quenching mechanism.

    \item We measure a remarkably small rest-UV size of $30\pm^{7}_{5}$ pc in the source plane for the galaxy. Implying a high stellar density, this hints at exhaustion of gas via rapid star formation and/or stellar feedback, and possibly the galaxy being a progenitor of small quiescent galaxies which have grown via dry merging.
    
    \item We can likely rule out environmental quenching and quenching due to AGN feedback, due to the early time and lack of broad emission in the galaxy. We cannot, however, rule out the scenario where MACS0417-z5PSB is a star-forming galaxy which is in a lull of a bursty star-formation history.

\end{enumerate}

More data is needed to fully investigate this source. To robustly determine the star formation rate via H$\alpha$ and the age of the stars via Balmer absorption lines, higher resolution rest-frame optical spectroscopy is needed. To investigate the amount of gas in the galaxy, signaling something about the mechanism which has ceased its star formation, observations with ALMA [CII] would be enlightening. If there is gas remaining in the galaxy, perhaps there would be a future of rejuvenation for MACS0417-z5PSB.

Finally, to understand the number density of nearly quiescent galaxies like MACS0417-z5PSB, population studies are necessary. In only a short time after the flight of JWST, several high-$z$ quiescent galaxies have been spectroscopically confirmed, a higher number than expected by models (e.g., \citealp{Schreiber2018, Cecchi2019, Girelli2019, Carnall2023b}). It is currently possible to identify galaxies lacking emission lines with NIRCam photometry alone, and it will be interesting to study number densities of quiescent galaxies at a variety of masses at $z>5$ in the coming months and years.


\begin{acknowledgments}
This research was enabled by grant 18JWST-GTO1 from the Canadian Space Agency and funding from the Natural Sciences and Engineering Research Council of Canada.
MB acknowledges support from the Slovenian national research agency ARRS through grant N1-0238 and the program HST-GO-16667, provided through a grant from the STScI under NASA contract NAS5-26555. This research used the Canadian Advanced Network For Astronomy Research (CANFAR) operated in partnership by the Canadian Astronomy Data Centre and The Digital Research Alliance of Canada with support from the National Research Council of Canada the Canadian Space Agency, CANARIE and the Canadian Foundation for Innovation. The Cosmic Dawn Center (DAWN) is funded by the Danish National Research Foundation under grant No. 140.
\end{acknowledgments}

%

\vspace{5mm}
\facilities{JWST (NIRSpec, NIRCam), HST (WFC3, ACS)}








\bibliography{library}{}

\begin{thebibliography}{}
\expandafter\ifx\csname natexlab\endcsname\relax\def\natexlab#1{#1}\fi
\providecommand{\url}[1]{\href{#1}{#1}}
\providecommand{\dodoi}[1]{doi:~\href{http://doi.org/#1}{\nolinkurl{#1}}}
\providecommand{\doeprint}[1]{\href{http://ascl.net/#1}{\nolinkurl{http://ascl.net/#1}}}
\providecommand{\doarXiv}[1]{\href{https://arxiv.org/abs/#1}{\nolinkurl{https://arxiv.org/abs/#1}}}

\bibitem[{{Almeida} {et~al.}(2023){Almeida}, {Nemmen}, \&
  {Riffel}}]{Almeida2023}
{Almeida}, I., {Nemmen}, R., \& {Riffel}, R. 2023, arXiv e-prints,
  arXiv:2303.00826, \dodoi{10.48550/arXiv.2303.00826}

\bibitem[{{Asada} {et~al.}(2022){Asada}, {Sawicki}, {Desprez}, {Abraham},
  {Brada{\v{c}}}, {Brammer}, {Harshan}, {Iyer}, {Martis}, {Mowla}, {Muzzin},
  {Noirot}, {Ravindranath}, {Sarrouh}, {Strait}, {Willott}, \&
  {Zabl}}]{Asada2022}
{Asada}, Y., {Sawicki}, M., {Desprez}, G., {et~al.} 2022, arXiv e-prints,
  arXiv:2212.07540, \dodoi{10.48550/arXiv.2212.07540}

\bibitem[{{Belli} {et~al.}(2019){Belli}, {Newman}, \& {Ellis}}]{Belli2019}
{Belli}, S., {Newman}, A.~B., \& {Ellis}, R.~S. 2019, \apj, 874, 17,
  \dodoi{10.3847/1538-4357/ab07af}

\bibitem[{{Bezanson} {et~al.}(2009){Bezanson}, {van Dokkum}, {Tal},
  {Marchesini}, {Kriek}, {Franx}, \& {Coppi}}]{Bezanson2009}
{Bezanson}, R., {van Dokkum}, P.~G., {Tal}, T., {et~al.} 2009, \apj, 697, 1290,
  \dodoi{10.1088/0004-637X/697/2/1290}

\bibitem[{{Carnall} {et~al.}(2018){Carnall}, {McLure}, {Dunlop}, \&
  {Dav{\'e}}}]{Carnall2018}
{Carnall}, A.~C., {McLure}, R.~J., {Dunlop}, J.~S., \& {Dav{\'e}}, R. 2018,
  \mnras, 480, 4379, \dodoi{10.1093/mnras/sty2169}

\bibitem[{{Carnall} {et~al.}(2023{\natexlab{a}}){Carnall}, {McLure}, {Dunlop},
  {McLeod}, {Wild}, {Cullen}, {Magee}, {Begley}, {Cimatti}, {Donnan},
  {Hamadouche}, {Jewell}, \& {Walker}}]{Carnall2023a}
{Carnall}, A.~C., {McLure}, R.~J., {Dunlop}, J.~S., {et~al.}
  2023{\natexlab{a}}, arXiv e-prints, arXiv:2301.11413,
  \dodoi{10.48550/arXiv.2301.11413}

\bibitem[{{Carnall} {et~al.}(2023{\natexlab{b}}){Carnall}, {McLeod}, {McLure},
  {Dunlop}, {Begley}, {Cullen}, {Donnan}, {Hamadouche}, {Jewell}, {Jones},
  {Pollock}, \& {Wild}}]{Carnall2023b}
{Carnall}, A.~C., {McLeod}, D.~J., {McLure}, R.~J., {et~al.}
  2023{\natexlab{b}}, \mnras, \dodoi{10.1093/mnras/stad369}

\bibitem[{{Cecchi} {et~al.}(2019){Cecchi}, {Bolzonella}, {Cimatti}, \&
  {Girelli}}]{Cecchi2019}
{Cecchi}, R., {Bolzonella}, M., {Cimatti}, A., \& {Girelli}, G. 2019, \apjl,
  880, L14, \dodoi{10.3847/2041-8213/ab2c80}

\bibitem[{{Charlot} \& {Fall}(2000)}]{Charlot2000}
{Charlot}, S., \& {Fall}, S.~M. 2000, \apj, 539, 718, \dodoi{10.1086/309250}

\bibitem[{{Ding} {et~al.}(2020){Ding}, {Silverman}, {Treu}, {Schulze},
  {Schramm}, {Birrer}, {Park}, {Jahnke}, {Bennert}, {Kartaltepe}, {Koekemoer},
  {Malkan}, \& {Sanders}}]{Ding2020}
{Ding}, X., {Silverman}, J., {Treu}, T., {et~al.} 2020, \apj, 888, 37,
  \dodoi{10.3847/1538-4357/ab5b90}

\bibitem[{{Donnari} {et~al.}(2021){Donnari}, {Pillepich}, {Nelson},
  {Marinacci}, {Vogelsberger}, \& {Hernquist}}]{Donnari2021}
{Donnari}, M., {Pillepich}, A., {Nelson}, D., {et~al.} 2021, \mnras, 506, 4760,
  \dodoi{10.1093/mnras/stab1950}

\bibitem[{{Fabian}(2012)}]{Fabian2012}
{Fabian}, A.~C. 2012, \araa, 50, 455,
  \dodoi{10.1146/annurev-astro-081811-125521}

\bibitem[{{Forrest} {et~al.}(2020){Forrest}, {Annunziatella}, {Wilson},
  {Marchesini}, {Muzzin}, {Cooper}, {Marsan}, {McConachie}, {Chan}, {Gomez},
  {Kado-Fong}, {L Barbera}, {Labb{\'e}}, {Lange-Vagle}, {Nantais}, {Nonino},
  {Pe{\~n}a}, {Saracco}, {Stefanon}, \& {van der Burg}}]{Forrest2020}
{Forrest}, B., {Annunziatella}, M., {Wilson}, G., {et~al.} 2020, \apjl, 890,
  L1, \dodoi{10.3847/2041-8213/ab5b9f}

\bibitem[{{Girelli} {et~al.}(2019){Girelli}, {Bolzonella}, \&
  {Cimatti}}]{Girelli2019}
{Girelli}, G., {Bolzonella}, M., \& {Cimatti}, A. 2019, \aap, 632, A80,
  \dodoi{10.1051/0004-6361/201834547}

\bibitem[{{Glazebrook} {et~al.}(2017){Glazebrook}, {Schreiber}, {Labb{\'e}},
  {Nanayakkara}, {Kacprzak}, {Oesch}, {Papovich}, {Spitler}, {Straatman},
  {Tran}, \& {Yuan}}]{Glazebrook2017}
{Glazebrook}, K., {Schreiber}, C., {Labb{\'e}}, I., {et~al.} 2017, \nat, 544,
  71, \dodoi{10.1038/nature21680}

\bibitem[{{Gould} {et~al.}(2023){Gould}, {Brammer}, {Valentino}, {Whitaker},
  {Weaver}, {Lagos}, {Rizzo}, {Franco}, {Hseih}, {Ilbert}, {Jin}, {Magdis},
  {McCracken}, {Mobasher}, {Shuntov}, {Steinhardt}, {Strait}, \&
  {Toft}}]{Gould2023}
{Gould}, K. M.~L., {Brammer}, G., {Valentino}, F., {et~al.} 2023, arXiv
  e-prints, arXiv:2302.10934, \dodoi{10.48550/arXiv.2302.10934}

\bibitem[{{Hopkins} {et~al.}(2010){Hopkins}, {Bundy}, {Hernquist}, {Wuyts}, \&
  {Cox}}]{Hopkins2010}
{Hopkins}, P.~F., {Bundy}, K., {Hernquist}, L., {Wuyts}, S., \& {Cox}, T.~J.
  2010, \mnras, 401, 1099, \dodoi{10.1111/j.1365-2966.2009.15699.x}

\bibitem[{{Horne}(1986)}]{Horne1986}
{Horne}, K. 1986, \pasp, 98, 609, \dodoi{10.1086/131801}

\bibitem[{{Iyer} {et~al.}(2020){Iyer}, {Tacchella}, {Genel}, {Hayward},
  {Hernquist}, {Brooks}, {Caplar}, {Dav{\'e}}, {Diemer}, {Forbes}, {Gawiser},
  {Somerville}, \& {Starkenburg}}]{Iyer2020}
{Iyer}, K.~G., {Tacchella}, S., {Genel}, S., {et~al.} 2020, \mnras, 498, 430,
  \dodoi{10.1093/mnras/staa2150}

\bibitem[{{Jullo} {et~al.}(2007){Jullo}, {Kneib}, {Limousin},
  {El{\'\i}asd{\'o}ttir}, {Marshall}, \& {Verdugo}}]{Jullo2007}
{Jullo}, E., {Kneib}, J.~P., {Limousin}, M., {et~al.} 2007, New Journal of
  Physics, 9, 447, \dodoi{10.1088/1367-2630/9/12/447}

\bibitem[{{King} \& {Pounds}(2015)}]{King2015}
{King}, A., \& {Pounds}, K. 2015, \araa, 53, 115,
  \dodoi{10.1146/annurev-astro-082214-122316}

\bibitem[{{Kneib} {et~al.}(1993){Kneib}, {Mellier}, {Fort}, \&
  {Mathez}}]{Kneib1993}
{Kneib}, J.~P., {Mellier}, Y., {Fort}, B., \& {Mathez}, G. 1993, \aap, 273, 367

\bibitem[{{Koudmani} {et~al.}(2022){Koudmani}, {Sijacki}, \&
  {Smith}}]{Koudmani2022}
{Koudmani}, S., {Sijacki}, D., \& {Smith}, M.~C. 2022, \mnras, 516, 2112,
  \dodoi{10.1093/mnras/stac2252}

\bibitem[{{Kroupa}(2001)}]{Kroupa2001}
{Kroupa}, P. 2001, \mnras, 322, 231, \dodoi{10.1046/j.1365-8711.2001.04022.x}

\bibitem[{{Legrand} {et~al.}(2022){Legrand}, {Hutter}, {Dayal}, {Ucci},
  {Gottl{\"o}ber}, \& {Yepes}}]{legrand2022}
{Legrand}, L., {Hutter}, A., {Dayal}, P., {et~al.} 2022, \mnras, 509, 595,
  \dodoi{10.1093/mnras/stab3034}

\bibitem[{{Looser} {et~al.}(2023){Looser}, {D'Eugenio}, {Maiolino}, {Witstok},
  {Sandles}, {Curtis-Lake}, {Chevallard}, {Tacchella}, {Johnson}, {Baker},
  {Suess}, {Carniani}, {Ferruit}, {Arribas}, {Bonaventura}, {Bunker},
  {Cameron}, {Charlot}, {Curti}, {de Graaff}, {Maseda}, {Rawle}, {Rix},
  {Rodriguez Del Pino}, {Smit}, {{\"U}bler}, {Willott}, {Alberts}, {Egami},
  {Eisenstein}, {Endsley}, {Hausen}, {Rieke}, {Robertson}, {Shivaei},
  {Williams}, {Boyett}, {Chen}, {Ji}, {Jones}, {Kumari}, {Nelson}, {Perna},
  {Saxena}, \& {Scholtz}}]{Looser2023}
{Looser}, T.~J., {D'Eugenio}, F., {Maiolino}, R., {et~al.} 2023, arXiv
  e-prints, arXiv:2302.14155.
\newblock \doarXiv{2302.14155}

\bibitem[{{Lovell} {et~al.}(2022){Lovell}, {Roper}, {Vijayan}, {Seeyave},
  {Irodotou}, {Wilkins}, {Conselice}, {Fortuni}, {Kuusisto}, {Merlin},
  {Santini}, \& {Thomas}}]{lovell2022}
{Lovell}, C.~C., {Roper}, W., {Vijayan}, A.~P., {et~al.} 2022, arXiv e-prints,
  arXiv:2211.07540, \dodoi{10.48550/arXiv.2211.07540}

\bibitem[{{Ma} {et~al.}(2019){Ma}, {Hayward}, {Casey}, {Hopkins}, {Quataert},
  {Liang}, {Faucher-Gigu{\`e}re}, {Feldmann}, \& {Kere{\v{s}}}}]{Ma2019}
{Ma}, X., {Hayward}, C.~C., {Casey}, C.~M., {et~al.} 2019, \mnras, 487, 1844,
  \dodoi{10.1093/mnras/stz1324}

\bibitem[{{Mahler} {et~al.}(2019){Mahler}, {Sharon}, {Fox}, {Coe}, {Jauzac},
  {Strait}, {Edge}, {Acebron}, {Andrade-Santos}, {Avila}, {Brada{\v{c}}},
  {Bradley}, {Carrasco}, {Cerny}, {Cibirka}, {Czakon}, {Dawson}, {Frye},
  {Hoag}, {Huang}, {Johnson}, {Jones}, {Kikuchihara}, {Lam}, {Livermore},
  {Lovisari}, {Mainali}, {Ogaz}, {Ouchi}, {Paterno-Mahler}, {Roederer}, {Ryan},
  {Salmon}, {Sendra-Server}, {Stark}, {Toft}, {Trenti}, {Umetsu}, {Vulcani}, \&
  {Zitrin}}]{Mahler2019}
{Mahler}, G., {Sharon}, K., {Fox}, C., {et~al.} 2019, \apj, 873, 96,
  \dodoi{10.3847/1538-4357/ab042b}

\bibitem[{{Man} \& {Belli}(2018)}]{Man2018}
{Man}, A., \& {Belli}, S. 2018, Nature Astronomy, 2, 695,
  \dodoi{10.1038/s41550-018-0558-1}

\bibitem[{{Manzano-King} {et~al.}(2019){Manzano-King}, {Canalizo}, \&
  {Sales}}]{ManzanoKing2019}
{Manzano-King}, C.~M., {Canalizo}, G., \& {Sales}, L.~V. 2019, \apj, 884, 54,
  \dodoi{10.3847/1538-4357/ab4197}

\bibitem[{{Marchesini} {et~al.}(2023){Marchesini}, {Brammer}, {Morishita},
  {Bergamini}, {Wang}, {Bradac}, {Roberts-Borsani}, {Strait}, {Treu},
  {Fontana}, {Jones}, {Santini}, {Vulcani}, {Acebron}, {Calabr{\`o}},
  {Castellano}, {Glazebrook}, {Grillo}, {Mercurio}, {Nanayakkara}, {Rosati},
  {Tubthong}, \& {Vanzella}}]{Marchesini2023}
{Marchesini}, D., {Brammer}, G., {Morishita}, T., {et~al.} 2023, \apjl, 942,
  L25, \dodoi{10.3847/2041-8213/acaaac}

\bibitem[{{McGee} {et~al.}(2014){McGee}, {Bower}, \& {Balogh}}]{McGee2014}
{McGee}, S.~L., {Bower}, R.~G., \& {Balogh}, M.~L. 2014, \mnras, 442, L105,
  \dodoi{10.1093/mnrasl/slu066}

\bibitem[{{Merlin} {et~al.}(2019){Merlin}, {Fortuni}, {Torelli}, {Santini},
  {Castellano}, {Fontana}, {Grazian}, {Pentericci}, {Pilo}, \&
  {Schmidt}}]{Merlin2019}
{Merlin}, E., {Fortuni}, F., {Torelli}, M., {et~al.} 2019, \mnras, 490, 3309,
  \dodoi{10.1093/mnras/stz2615}

\bibitem[{{Morishita} {et~al.}(2022){Morishita}, {Roberts-Borsani}, {Treu},
  {Brammer}, {Mason}, {Trenti}, {Vulcani}, {Wang}, {Acebron}, {Bah{\'e}},
  {Bergamini}, {Boyett}, {Bradac}, {Calabr{\`o}}, {Castellano}, {Chen}, {De
  Lucia}, {Filippenko}, {Fontana}, {Glazebrook}, {Grillo}, {Henry}, {Jones},
  {Kelly}, {Koekemoer}, {Leethochawalit}, {Lu}, {Marchesini}, {Mascia},
  {Mercurio}, {Merlin}, {Metha}, {Nanayakkara}, {Nonino}, {Paris},
  {Pentericci}, {Santini}, {Strait}, {Vanzella}, {Windhorst}, {Rosati}, \&
  {Xie}}]{Morishita2022b}
{Morishita}, T., {Roberts-Borsani}, G., {Treu}, T., {et~al.} 2022, arXiv
  e-prints, arXiv:2211.09097, \dodoi{10.48550/arXiv.2211.09097}

\bibitem[{{Naab} \& {Ostriker}(2009)}]{Naab2009}
{Naab}, T., \& {Ostriker}, J.~P. 2009, \apj, 690, 1452,
  \dodoi{10.1088/0004-637X/690/2/1452}

\bibitem[{{Nanayakkara} {et~al.}(2022){Nanayakkara}, {Glazebrook}, {Jacobs},
  {Schreiber}, {Brammer}, {Esdaile}, {Kacprzak}, {Labbe}, {Lagos},
  {Marchesini}, {Marsan}, {Nateghi}, {Oesch}, {Papovich}, {Remus}, \&
  {Tran}}]{Nanayakkara2022}
{Nanayakkara}, T., {Glazebrook}, K., {Jacobs}, C., {et~al.} 2022, arXiv
  e-prints, arXiv:2212.11638, \dodoi{10.48550/arXiv.2212.11638}

\bibitem[{{Nelson} {et~al.}(2018){Nelson}, {Pillepich}, {Springel},
  {Weinberger}, {Hernquist}, {Pakmor}, {Genel}, {Torrey}, {Vogelsberger},
  {Kauffmann}, {Marinacci}, \& {Naiman}}]{Nelson2018}
{Nelson}, D., {Pillepich}, A., {Springel}, V., {et~al.} 2018, \mnras, 475, 624,
  \dodoi{10.1093/mnras/stx3040}

\bibitem[{{Nelson} {et~al.}(2019){Nelson}, {Pillepich}, {Springel}, {Pakmor},
  {Weinberger}, {Genel}, {Torrey}, {Vogelsberger}, {Marinacci}, \&
  {Hernquist}}]{Nelson2019}
---. 2019, \mnras, 490, 3234, \dodoi{10.1093/mnras/stz2306}

\bibitem[{{Noirot} {et~al.}(2022{\natexlab{a}}){Noirot}, {Sawicki}, {Abraham},
  {Brada{\v{c}}}, {Iyer}, {Moutard}, {Pacifici}, {Ravindranath}, \&
  {Willott}}]{Noirot2022a}
{Noirot}, G., {Sawicki}, M., {Abraham}, R., {et~al.} 2022{\natexlab{a}},
  \mnras, 512, 3566, \dodoi{10.1093/mnras/stac668}

\bibitem[{{Noirot} {et~al.}(2022{\natexlab{b}}){Noirot}, {Desprez}, {Asada},
  {Sawicki}, {Estrada-Carpenter}, {Martis}, {Sarrouh}, {Strait}, {Abraham},
  {Brada{\v{c}}}, {Brammer}, {Iyer}, {MacFarland}, {Matharu}, {Mowla},
  {Muzzin}, {Pacifici}, {Ravindranath}, {Willott}, {Albert}, {Doyon},
  {Hutchings}, \& {Rowlands}}]{Noirot2022b}
{Noirot}, G., {Desprez}, G., {Asada}, Y., {et~al.} 2022{\natexlab{b}}, arXiv
  e-prints, arXiv:2212.07366, \dodoi{10.48550/arXiv.2212.07366}

\bibitem[{{Ono} {et~al.}(2022){Ono}, {Harikane}, {Ouchi}, {Yajima}, {Abe},
  {Isobe}, {Shibuya}, {Zhang}, {Nakajima}, \& {Umeda}}]{Ono2022}
{Ono}, Y., {Harikane}, Y., {Ouchi}, M., {et~al.} 2022, arXiv e-prints,
  arXiv:2208.13582, \dodoi{10.48550/arXiv.2208.13582}

\bibitem[{{Peng} {et~al.}(2010){Peng}, {Lilly}, {Kova{\v{c}}}, {Bolzonella},
  {Pozzetti}, {Renzini}, {Zamorani}, {Ilbert}, {Knobel}, {Iovino}, {Maier},
  {Cucciati}, {Tasca}, {Carollo}, {Silverman}, {Kampczyk}, {de Ravel},
  {Sanders}, {Scoville}, {Contini}, {Mainieri}, {Scodeggio}, {Kneib}, {Le
  F{\`e}vre}, {Bardelli}, {Bongiorno}, {Caputi}, {Coppa}, {de la Torre},
  {Franzetti}, {Garilli}, {Lamareille}, {Le Borgne}, {Le Brun}, {Mignoli},
  {Perez Montero}, {Pello}, {Ricciardelli}, {Tanaka}, {Tresse}, {Vergani},
  {Welikala}, {Zucca}, {Oesch}, {Abbas}, {Barnes}, {Bordoloi}, {Bottini},
  {Cappi}, {Cassata}, {Cimatti}, {Fumana}, {Hasinger}, {Koekemoer},
  {Leauthaud}, {Maccagni}, {Marinoni}, {McCracken}, {Memeo}, {Meneux}, {Nair},
  {Porciani}, {Presotto}, \& {Scaramella}}]{Peng2010}
{Peng}, Y.-j., {Lilly}, S.~J., {Kova{\v{c}}}, K., {et~al.} 2010, \apj, 721,
  193, \dodoi{10.1088/0004-637X/721/1/193}

\bibitem[{{Reines}(2022)}]{Reines2022}
{Reines}, A.~E. 2022, Nature Astronomy, 6, 26,
  \dodoi{10.1038/s41550-021-01556-0}

\bibitem[{{Salmon} {et~al.}(2015){Salmon}, {Papovich}, {Finkelstein}, {Tilvi},
  {Finlator}, {Behroozi}, {Dahlen}, {Dav{\'e}}, {Dekel}, {Dickinson},
  {Ferguson}, {Giavalisco}, {Long}, {Lu}, {Mobasher}, {Reddy}, {Somerville}, \&
  {Wechsler}}]{Salmon2015}
{Salmon}, B., {Papovich}, C., {Finkelstein}, S.~L., {et~al.} 2015, \apj, 799,
  183, \dodoi{10.1088/0004-637X/799/2/183}

\bibitem[{{Santini} {et~al.}(2021){Santini}, {Castellano}, {Merlin}, {Fontana},
  {Fortuni}, {Kodra}, {Magnelli}, {Menci}, {Calabr{\`o}}, {Lovell},
  {Pentericci}, {Testa}, \& {Wilkins}}]{Santini2021}
{Santini}, P., {Castellano}, M., {Merlin}, E., {et~al.} 2021, \aap, 652, A30,
  \dodoi{10.1051/0004-6361/202039738}

\bibitem[{{Santini} {et~al.}(2022){Santini}, {Castellano}, {Fontana},
  {Fortuni}, {Menci}, {Merlin}, {Pagul}, {Testa}, {Calabr{\`o}}, {Paris}, \&
  {Pentericci}}]{Santini2022}
{Santini}, P., {Castellano}, M., {Fontana}, A., {et~al.} 2022, \apj, 940, 135,
  \dodoi{10.3847/1538-4357/ac9a48}

\bibitem[{{Schreiber} {et~al.}(2018){Schreiber}, {Glazebrook}, {Nanayakkara},
  {Kacprzak}, {Labb{\'e}}, {Oesch}, {Yuan}, {Tran}, {Papovich}, {Spitler}, \&
  {Straatman}}]{Schreiber2018}
{Schreiber}, C., {Glazebrook}, K., {Nanayakkara}, T., {et~al.} 2018, \aap, 618,
  A85, \dodoi{10.1051/0004-6361/201833070}

\bibitem[{{Sharma} {et~al.}(2022){Sharma}, {Brooks}, {Tremmel}, {Bellovary}, \&
  {Quinn}}]{Sharma2022}
{Sharma}, R.~S., {Brooks}, A.~M., {Tremmel}, M., {Bellovary}, J., \& {Quinn},
  T.~R. 2022, arXiv e-prints, arXiv:2211.05275,
  \dodoi{10.48550/arXiv.2211.05275}

\bibitem[{{Sherman} {et~al.}(2020){Sherman}, {Jogee}, {Florez}, {Stevans},
  {Kawinwanichakij}, {Wold}, {Finkelstein}, {Papovich}, {Ciardullo},
  {Gronwall}, {Cora}, {Hough}, \& {Vega-Mart{\'\i}nez}}]{Sherman2020}
{Sherman}, S., {Jogee}, S., {Florez}, J., {et~al.} 2020, \mnras, 499, 4239,
  \dodoi{10.1093/mnras/staa3167}

\bibitem[{{Tacchella} {et~al.}(2016){Tacchella}, {Dekel}, {Carollo},
  {Ceverino}, {DeGraf}, {Lapiner}, {Mandelker}, \& {Primack}}]{Tacchella2016}
{Tacchella}, S., {Dekel}, A., {Carollo}, C.~M., {et~al.} 2016, \mnras, 458,
  242, \dodoi{10.1093/mnras/stw303}

\bibitem[{{Tacchella} {et~al.}(2022){Tacchella}, {Conroy}, {Faber}, {Johnson},
  {Leja}, {Barro}, {Cunningham}, {Deason}, {Guhathakurta}, {Guo}, {Hernquist},
  {Koo}, {McKinnon}, {Rockosi}, {Speagle}, {van Dokkum}, \&
  {Yesuf}}]{Tacchella2022}
{Tacchella}, S., {Conroy}, C., {Faber}, S.~M., {et~al.} 2022, \apj, 926, 134,
  \dodoi{10.3847/1538-4357/ac449b}

\bibitem[{{Trayford} {et~al.}(2015){Trayford}, {Theuns}, {Bower}, {Schaye},
  {Furlong}, {Schaller}, {Frenk}, {Crain}, {Dalla Vecchia}, \&
  {McCarthy}}]{Trayford2015}
{Trayford}, J.~W., {Theuns}, T., {Bower}, R.~G., {et~al.} 2015, \mnras, 452,
  2879, \dodoi{10.1093/mnras/stv1461}

\bibitem[{{Valentino} {et~al.}(2020){Valentino}, {Tanaka}, {Davidzon}, {Toft},
  {G{\'o}mez-Guijarro}, {Stockmann}, {Onodera}, {Brammer}, {Ceverino},
  {Faisst}, {Gallazzi}, {Hayward}, {Ilbert}, {Kubo}, {Magdis}, {Selsing},
  {Shimakawa}, {Sparre}, {Steinhardt}, {Yabe}, \& {Zabl}}]{Valentino2020}
{Valentino}, F., {Tanaka}, M., {Davidzon}, I., {et~al.} 2020, \apj, 889, 93,
  \dodoi{10.3847/1538-4357/ab64dc}

\bibitem[{{van der Wel} {et~al.}(2014){van der Wel}, {Franx}, {van Dokkum},
  {Skelton}, {Momcheva}, {Whitaker}, {Brammer}, {Bell}, {Rix}, {Wuyts},
  {Ferguson}, {Holden}, {Barro}, {Koekemoer}, {Chang}, {McGrath},
  {H{\"a}ussler}, {Dekel}, {Behroozi}, {Fumagalli}, {Leja}, {Lundgren},
  {Maseda}, {Nelson}, {Wake}, {Patel}, {Labb{\'e}}, {Faber}, {Grogin}, \&
  {Kocevski}}]{vanderWel2014}
{van der Wel}, A., {Franx}, M., {van Dokkum}, P.~G., {et~al.} 2014, \apj, 788,
  28, \dodoi{10.1088/0004-637X/788/1/28}

\bibitem[{{Ward} {et~al.}(2022){Ward}, {Harrison}, {Costa}, \&
  {Mainieri}}]{Ward2022}
{Ward}, S.~R., {Harrison}, C.~M., {Costa}, T., \& {Mainieri}, V. 2022, \mnras,
  514, 2936, \dodoi{10.1093/mnras/stac1219}

\bibitem[{{Weaver} {et~al.}(2022){Weaver}, {Davidzon}, {Toft}, {Ilbert},
  {McCracken}, {Gould}, {Jespersen}, {Steinhardt}, {Lagos}, {Capak}, {Casey},
  {Chartab}, {Faisst}, {Hayward}, {Kartaltepe}, {Kauffmann}, {Koekemoer},
  {Kokorev}, {Laigle}, {Liu}, {Long}, {Magdis}, {McPartland}, {Milvang-Jensen},
  {Mobasher}, {Moneti}, {Peng}, {Sanders}, {Shuntov}, {Sneppen}, {Valentino},
  {Zalesky}, \& {Zamorani}}]{Weaver2022}
{Weaver}, J.~R., {Davidzon}, I., {Toft}, S., {et~al.} 2022, arXiv e-prints,
  arXiv:2212.02512, \dodoi{10.48550/arXiv.2212.02512}

\bibitem[{{Wellons} {et~al.}(2023){Wellons}, {Faucher-Gigu{\`e}re}, {Hopkins},
  {Quataert}, {Angl{\'e}s-Alc{\'a}zar}, {Feldmann}, {Hayward}, {Kere{\v{s}}},
  {Su}, \& {Wetzel}}]{Wellons2023}
{Wellons}, S., {Faucher-Gigu{\`e}re}, C.-A., {Hopkins}, P.~F., {et~al.} 2023,
  \mnras, 520, 5394, \dodoi{10.1093/mnras/stad511}

\bibitem[{{Whitaker} {et~al.}(2021){Whitaker}, {Williams}, {Mowla}, {Spilker},
  {Toft}, {Narayanan}, {Pope}, {Magdis}, {van Dokkum}, {Akhshik}, {Bezanson},
  {Brammer}, {Leja}, {Man}, {Nelson}, {Richard}, {Pacifici}, {Sharon}, \&
  {Valentino}}]{Whitaker2021}
{Whitaker}, K.~E., {Williams}, C.~C., {Mowla}, L., {et~al.} 2021, \nat, 597,
  485, \dodoi{10.1038/s41586-021-03806-7}

\bibitem[{{Williams} {et~al.}(2021){Williams}, {Spilker}, {Whitaker},
  {Dav{\'e}}, {Woodrum}, {Brammer}, {Bezanson}, {Narayanan}, \&
  {Weiner}}]{Williams2021}
{Williams}, C.~C., {Spilker}, J.~S., {Whitaker}, K.~E., {et~al.} 2021, \apj,
  908, 54, \dodoi{10.3847/1538-4357/abcbf6}

\bibitem[{{Williams} {et~al.}(2022){Williams}, {Kelly}, {Chen}, {Brammer},
  {Zitrin}, {Treu}, {Scarlata}, {Koekemoer}, {Oguri}, {Lin}, {Diego}, {Nonino},
  {Hjorth}, {Langeroodi}, {Broadhurst}, {Rogers}, {Perez-Fournon}, {Foley},
  {Jha}, {Filippenko}, {Strolger}, {Pierel}, {Poidevin}, \&
  {Yang}}]{Williams2022}
{Williams}, H., {Kelly}, P.~L., {Chen}, W., {et~al.} 2022, arXiv e-prints,
  arXiv:2210.15699, \dodoi{10.48550/arXiv.2210.15699}

\bibitem[{{Willott} {et~al.}(2022){Willott}, {Doyon}, {Albert}, {Brammer},
  {Dixon}, {Muzic}, {Ravindranath}, {Scholz}, {Abraham}, {Artigau},
  {Brada{\v{c}}}, {Goudfrooij}, {Hutchings}, {Iyer}, {Jayawardhana}, {LaMassa},
  {Martis}, {Meyer}, {Morishita}, {Mowla}, {Muzzin}, {Noirot}, {Pacifici},
  {Rowlands}, {Sarrouh}, {Sawicki}, {Taylor}, {Volk}, \& {Zabl}}]{Willott2022}
{Willott}, C.~J., {Doyon}, R., {Albert}, L., {et~al.} 2022, \pasp, 134, 025002,
  \dodoi{10.1088/1538-3873/ac5158}

\bibitem[{{Wu} {et~al.}(2018){Wu}, {van der Wel}, {Bezanson}, {Gallazzi},
  {Pacifici}, {Straatman}, {Bari{\v{s}}i{\'c}}, {Bell}, {Chauke}, {van Houdt},
  {Franx}, {Muzzin}, {Sobral}, \& {Wild}}]{Wu2018}
{Wu}, P.-F., {van der Wel}, A., {Bezanson}, R., {et~al.} 2018, \apj, 868, 37,
  \dodoi{10.3847/1538-4357/aae822}

\bibitem[{{Wu} {et~al.}(2021){Wu}, {Nelson}, {van der Wel}, {Pillepich},
  {Zibetti}, {Bezanson}, {DEugenio}, {Gallazzi}, {Pacifici}, {Straatman},
  {Bari{\v{s}}i{\'c}}, {Bell}, {Maseda}, {Muzzin}, {Sobral}, \&
  {Whitaker}}]{Wu2021}
{Wu}, P.-F., {Nelson}, D., {van der Wel}, A., {et~al.} 2021, \aj, 162, 201,
  \dodoi{10.3847/1538-3881/ac20d6}

\bibitem[{{Wuyts} {et~al.}(2013){Wuyts}, {F{\"o}rster Schreiber}, {Nelson},
  {van Dokkum}, {Brammer}, {Chang}, {Faber}, {Ferguson}, {Franx}, {Fumagalli},
  {Genzel}, {Grogin}, {Kocevski}, {Koekemoer}, {Lundgren}, {Lutz}, {McGrath},
  {Momcheva}, {Rosario}, {Skelton}, {Tacconi}, {van der Wel}, \&
  {Whitaker}}]{Wuyts2013}
{Wuyts}, S., {F{\"o}rster Schreiber}, N.~M., {Nelson}, E.~J., {et~al.} 2013,
  \apj, 779, 135, \dodoi{10.1088/0004-637X/779/2/135}

\bibitem[{{Yang} {et~al.}(2020){Yang}, {Birrer}, \& {Treu}}]{Yang2020}
{Yang}, L., {Birrer}, S., \& {Treu}, T. 2020, \mnras, 496, 2648,
  \dodoi{10.1093/mnras/staa1649}

\bibitem[{{Yang} {et~al.}(2022){Yang}, {Morishita}, {Leethochawalit},
  {Castellano}, {Calabr{\`o}}, {Treu}, {Bonchi}, {Fontana}, {Mason}, {Merlin},
  {Paris}, {Trenti}, {Roberts-Borsani}, {Bradac}, {Vanzella}, {Vulcani},
  {Marchesini}, {Ding}, {Nanayakkara}, {Birrer}, {Glazebrook}, {Jones},
  {Boyett}, {Santini}, {Strait}, \& {Wang}}]{Yang2022}
{Yang}, L., {Morishita}, T., {Leethochawalit}, N., {et~al.} 2022, \apjl, 938,
  L17, \dodoi{10.3847/2041-8213/ac8803}

\bibitem[{{Zolotov} {et~al.}(2015){Zolotov}, {Dekel}, {Mandelker}, {Tweed},
  {Inoue}, {DeGraf}, {Ceverino}, {Primack}, {Barro}, \& {Faber}}]{Zolotov2015}
{Zolotov}, A., {Dekel}, A., {Mandelker}, N., {et~al.} 2015, \mnras, 450, 2327,
  \dodoi{10.1093/mnras/stv740}

\end{thebibliography}
\bibliographystyle{aasjournal}

\end{document}